\documentclass[onecolumn,showpacs,preprintnumbers,amsmath,amssymb]{revtex4}

\usepackage{graphicx}
\usepackage{epsfig}

\def\dd{\displaystyle}

\begin{document}
\title{\bf The Dirichlet Casimir Energy For $\phi^4$ Theory in a Rectangular Waveguide}

\author{M. A. Valuyan}
\email{m-valuyan@sbu.ac.ir; m.valuyan@semnaniau.ac.ir}
\affiliation{Department of Physics, Semnan Branch, Islamic Azad University, Semnan, Iran}
\date{\today}

\begin{abstract}
In this paper, we presented the zero- and first-order radiative corrections to the Casimir energy for a massive scalar field confined with Dirichlet boundary condition in an open-ended rectangular waveguide. In the calculation procedure, we applied a systematic renormalization program that allows all influences imposed by dominant boundary conditions in a problem be automatically reflected in the counterterms, leading the counterterms to be obtained in a position-dependent manner. To remove the appeared divergences in the computation task, the Box Subtraction Scheme as a regularization technique was used. In this regularization technique, usually, two similar configurations were introduced. Then, to find the Casimir energy, the zero point energies of these two configurations were subtracted from each other via defining appropriate limits. In the present work, first, the leading-order Casimir energy for the massive scalar field in a waveguide is briefly presented. Next, by applying this renormalization and regularization procedures, the first-order radiative correction to the Casimir energy in the waveguide is calculated. Finally, all the necessary limits of the obtained answers for massive and massless cases are computed and the consistency of the obtained results are discussed.
\end{abstract}

\maketitle

\section{Introduction}
\label{sec:intro}
Discussing the attractive force between two parallel plates has a long history. One of the most important parts of this history is the Casimir's prediction. In his famous essay, Casimir explained the attraction between two uncharged perfectly conducting parallel plates due to the polarization of electromagnetic field\,\cite{h.b.g.}. This prediction, which was firstly examined in 1958 by Sparnaay\,\cite{Sparnaay.M.J.}, later received growing attention, leading to several applications in many fields of physics\,\cite{quantum.field.theory.1,condensed.matter.1,atomic.molecular.1,astro.physics.1,Mathematical.physics.1}. The Casimir energy and its related force have been investigated for several known quantum fields and configurations with multiple boundary conditions. Additionally, this energy and its related force for interacting quantum field theory have been extensively studied. The first attempt to calculate the radiative correction to the Casimir energy of electromagnetic field was conducted by Bordag et al.\,\cite{Bordag.et.al,bordag.linding.}. Later, several attempts were made in a radiative correction to the Casimir energy for other quantum fields with multiple boundary conditions\,\cite{radiative.correction.free.counterterms.1,radiative.correction.free.counterterms.2,radiative.correction.free.counterterms.3, radiative.correction.free.counterterms.4,Robaschik.}. In this category of problems, a renormalization program is typically used to renormalize the bare parameters. In the most of the previous works, to renormalize the bare parameters\,(\emph{e.g.}, the mass of the field and coupling constant), the counterterm related to the free theory is imported in the calculation procedure. These counterterms have also been used even for bounded fields in the presence of boundary conditions. It is expected that when non-trivial boundary condition or topology influences the quantum field, this non-triviality to be reflected in all elements of the renormalization program including the counterterm. Moreover, since counterterms must renormalize the bare parameters of the problem, if they are not chosen properly, all divergences will not be removed correctly. Thus, some physical quantities ultimately take divergent values\,\cite{cavalcanti.}. To resolve this problem, a new way for renormalization of the bare parameters was prescribed in Refs.\,\cite{posision.dependent.counterterms.works.ours.1,posision.dependent.counterterms.works.ours.2,posision.dependent.counterterms.works.others.1, posision.dependent.counterterms.works.others.2,posision.dependent.counterterms.works.others.3}. In their renormalization program, all influences of boundary conditions were imported in the counterterms using the Green's function. This point has caused the counterterms to be obtained in a position-dependent manner. In the present study, assuming the correctness of their hypothesis, we allow the counterterms be extracted automatically from the renormalization program. This made the obtained counterterms to be position-dependent. Using this counterterm, the vacuum energy of our system was calculated systematically up to the first order of coupling constant $\lambda$. This renormalization program was successful and its final solution was consistent with known physical principles.
\par
\begin{figure}[th] \hspace{0cm}\includegraphics[width=5cm]{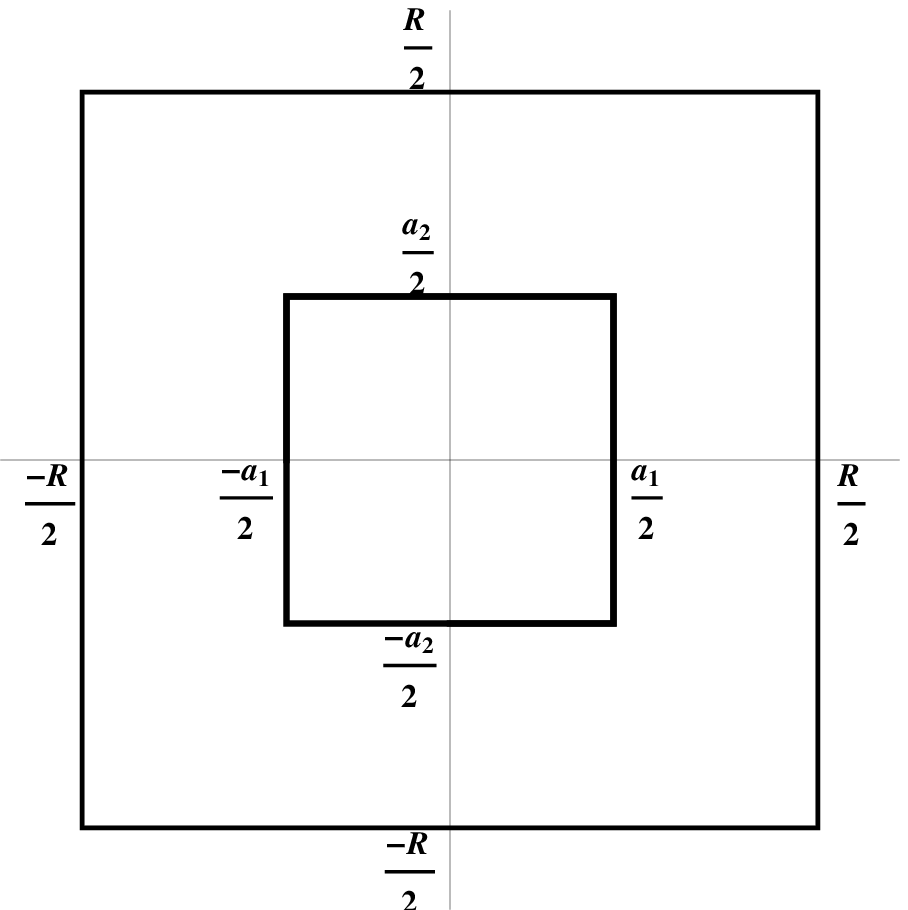}\hspace{0.5cm}\includegraphics[width=5cm]{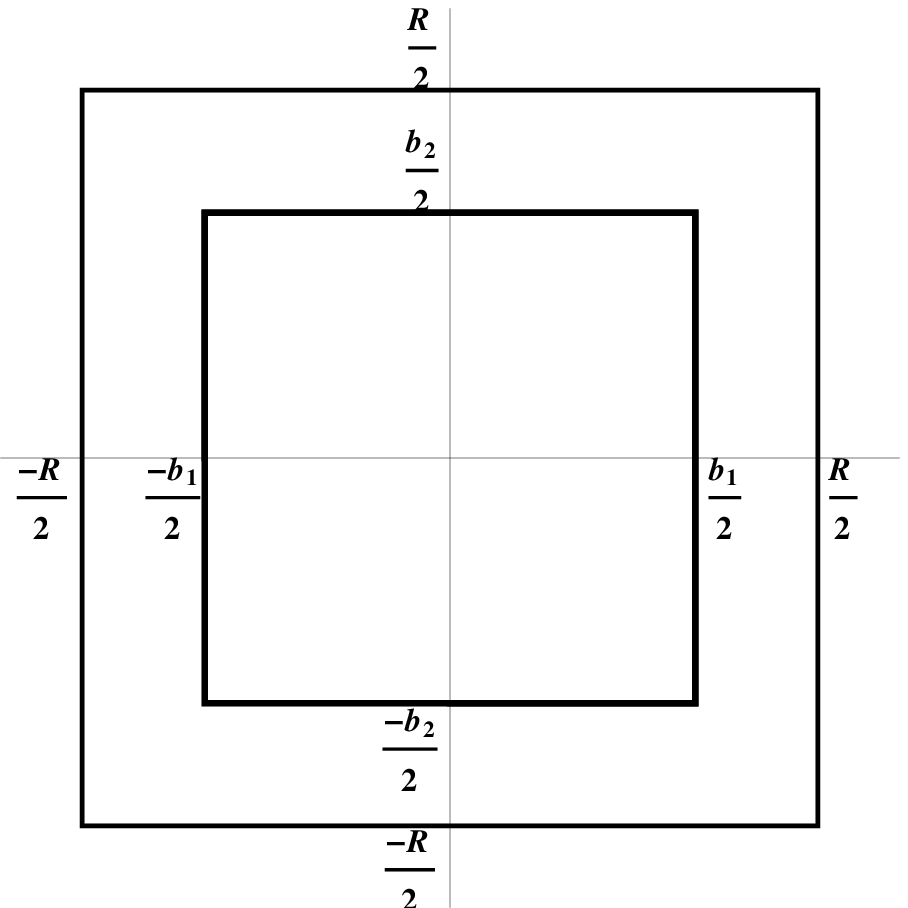}\caption{\label{fig.1} \small
  The Left figure is ``$A$ configuration" and the right one is ``$B$
  configuration".}
  \label{geometry.1}
\end{figure}

In the common definition of the Casimir energy, the contribution of two vacuum energies is subtracted from each other. Since the vacuum energy has an infinite value, the need for using one or more regularization techniques is inevitable. Zeta function regularization, Green's function method, heat kernel series, and Box Subtraction Scheme\,(BSS) are some of the important known regularization techniques in this context\,\cite{Edery.,regularization.methods.1.,regularization.methods.2.,regularization.methods.3.,regularization.methods.4.,regularization.methods.5.}.
In the present work, we used the BSS that is a slight modification of Boyer's method\,\cite{boyer.}. In this method, two similar configurations are introduced and then the vacuum energies of these two configurations are subtracted from each other in proper limits\,\cite{BSS.2,BSS.3,BSS.4}. This subtraction is equivalent to the work done in deforming of configurations and thus the result per unit volume is expected to be finite on physical grounds. Also, defining two similar configurations provides a situation that allows importing more parameters in the calculation procedure. These added parameters usually play a regulatory role and provide sufficient degrees of freedom in the divergence removal. These parameters also play an important role in clarifying the process of eliminating the divergences. This method significantly reduces the need for using the analytic continuation techniques that generate numerous possible ambiguities in the calculation of the Casimir energy. As mentioned earlier, in the definition of the BSS we need to introduce two similar configurations. Therefore, to calculate the Casimir energy for an infinite rectangular waveguide with a cross-section $a_1\times a_2$\,(Fig.\,(\ref{fig.1})), we introduce two similar waveguides. Fig.\,(\ref{fig.1}) presents the cross-section of two different rectangular open-ended waveguides trapped in a larger waveguide with a cross-section of $R\times R$. The Casimir energy can now be defined as:
\begin{equation}\label{BSS.CAS.Def.}
  E_{{\rm{Cas.}}}  = \mathop {\lim }\limits_{b_1 /a,b_2 /a \to
  \infty } \left[ {\mathop {\lim }\limits_{R/b \to \infty } \left(
  {E_{A}- E_{B} } \right)} \right],
\end{equation}
\begin{figure}[th] \hspace{0cm}\includegraphics[width=6cm]{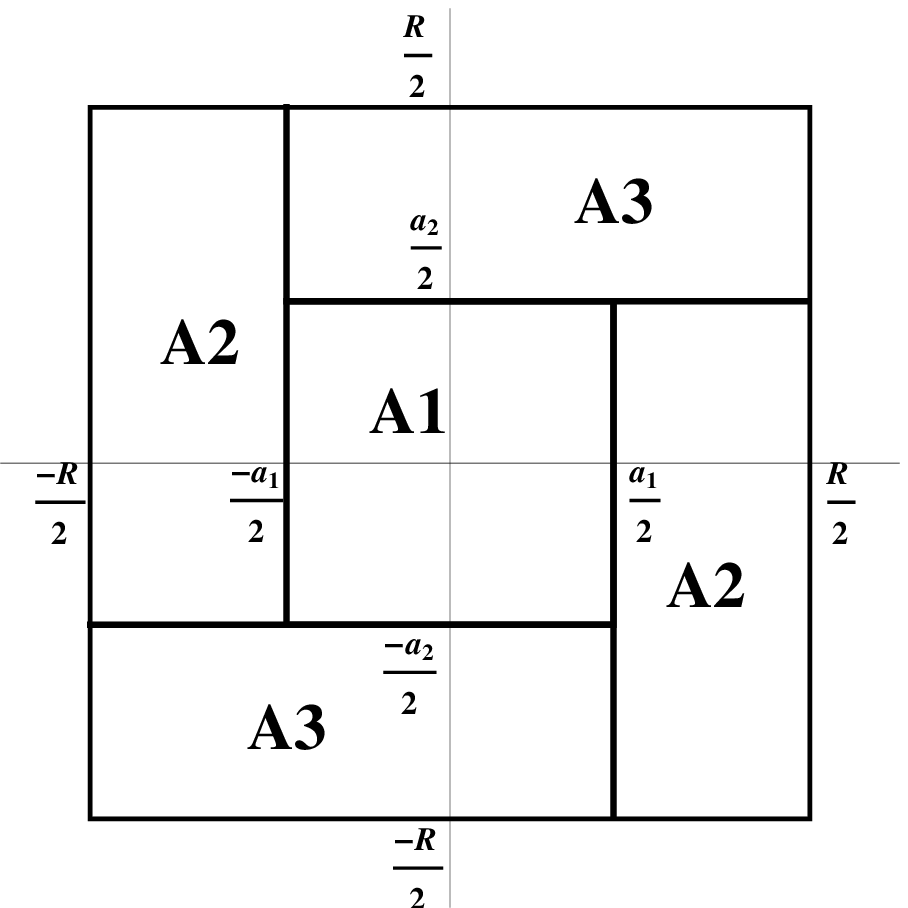}\hspace{0.5cm}\includegraphics[width=6cm]{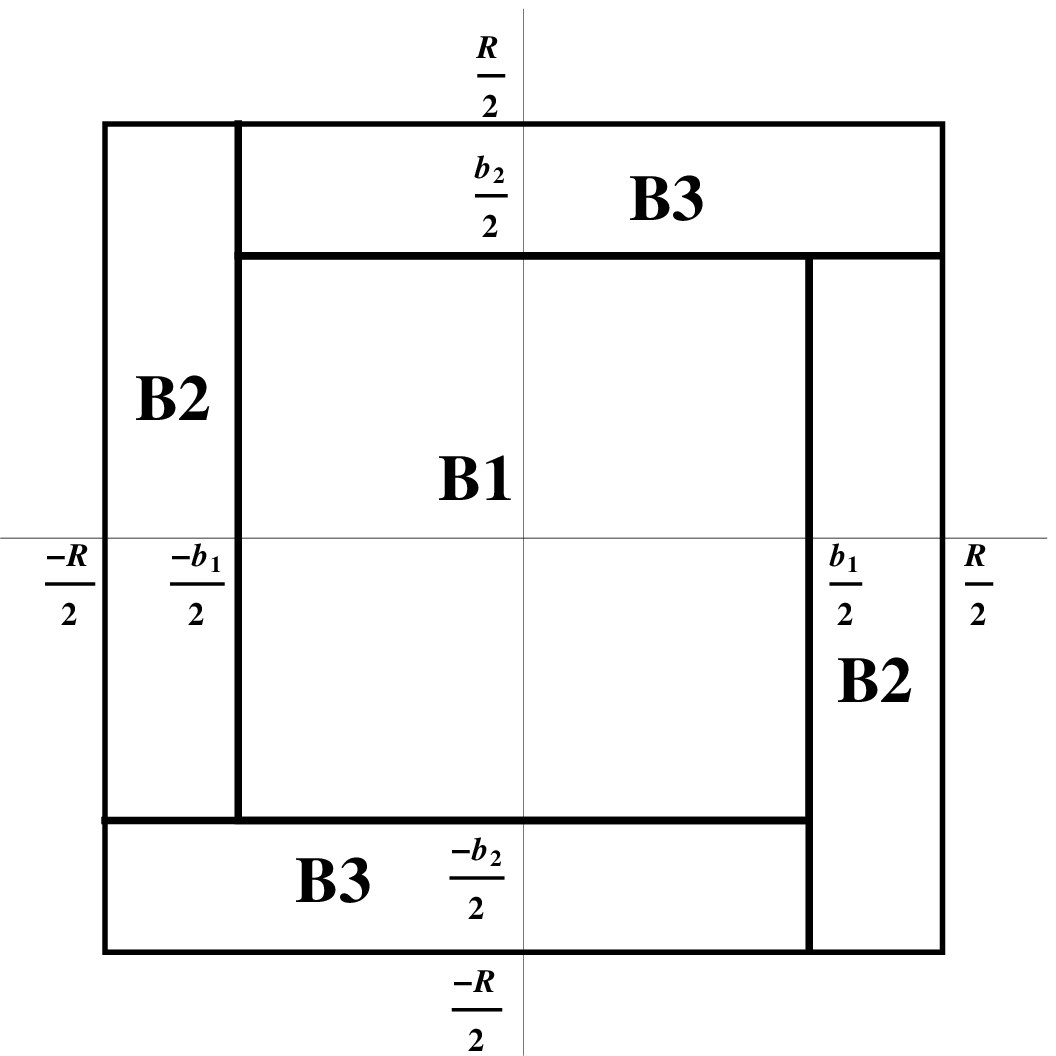}\caption{\label{fig.2} \small
  The Left figure is ``$A'$ configuration" and the right one is ``$B'$
  configuration". To calculate the Casimir energy, the zero-point energies of these two configurations should be subtracted according to Eq.\,\eqref{BSS.CAS.Def.}. In the final step, the size of configuration B goes to infinity, while the other parameters of the problem are kept constant.}
  \label{geometry.2}
\end{figure}
where $E_A$ ($E_B$) is the vacuum energy of configuration A (B), $a\equiv \mbox{Max}\{a_{1},a_{2}\}$, and $b\equiv\mbox{Max}\{b_{1},b_{2}\}$. To calculate the Casimir energy, it is necessary to estimate the vacuum energies in the whole configuration. However, calculation of this quantity in the middle region of defined configurations is a cumbersome task. Therefore, to simplify the task, we defined an alternative set of configurations in Fig.\,(\ref{fig.2}). We can then define the Casimir energy as in Eq.\,\eqref{BSS.CAS.Def.}, but with following replacements $A\rightarrow A'$
and $B\rightarrow B'$. In configurations $A'$ and $B'$ shown in Fig.\,(\ref{fig.2}), the middle region is divided into four waveguides, which are placed around the central waveguide. Therefore, the subtraction of vacuum energies for the new set of configurations can be written as:
\begin{eqnarray}\label{subtraction.vacuums.}
 \Delta E_{\mbox{\tiny Vac.}}= E_{A'}-E_{B'}=E_{A1}+2E_{A2}+2E_{A3}-E_{B1}-2E_{B2}-2E_{B3},
\end{eqnarray}
where $E_{A1}$, $E_{A2}$, $E_{A3}$, $E_{B1}$, $E_{B2}$, and $E_{B3}$ are the vacuum energies of corresponding regions. The configurations displayed in Fig.\,(\ref{fig.2}), which was introduced previously in\,\cite{BSS.1}, were successfully applied for calculation of the electromagnetic Casimir energy. Adding extra lines in the outer region of Fig.\,(\ref{fig.1}), may raise the question of whether these lines affect the Casimir energy. To deal with this concern, one can note that the remaining finite contribution to the Casimir energy coming from the outer waveguides, even after the BSS, is nonzero for finite values of the dimensions of the waveguides. However, as shown in Appendix\,\ref{appendix.outer.region}, in the large range of $R$, there is a partial cancellation between those terms, and the remaining terms tend to zero in the limit $R\to\infty$. Therefore, it can be stated that the outer waveguides have done their job in the BSS for canceling of infinities without leaving any finite contribution to the Casimir energy in the limit $R\to\infty$. The calculation presented in Appendix\,\ref{appendix.outer.region} proves that the outer waveguides and their remaining boundaries do not affect the Casimir energy of the original waveguide\,(Region $A1$). It should be noted that a similar proof to this one was made previously for calculating the electromagnetic Casimir energy inside a conducting rectangular waveguide and the same results were obtained\,(see Appendix B in\,\cite{BSS.1}). In this study, by applying the second sets of configurations shown in Fig.\,(\ref{fig.2}) and the mentioned renormalization program, the zero- and first-order radiative corrections were computed on the Casimir energy for the massive scalar field in $\phi^4$ theory with Dirichlet boundary condition in an open-ended rectangular waveguide. This problem in the case of first-order radiative correction to the Casimir energy is novel. Additionally, performing this renormalization program supplemented by the BSS creates an exclusive method that does not need any analytic continuation technique. Therefore, all possible ambiguities caused by the analytic continuation techniques are overcome in the calculation process. Eventually, we validated the correctness of the obtained results by checking the consistency with previously analyzed settings in specific limit cases.
\par
The leading order Casimir energy in a rectangular cavity with multiple types of fields and boundary conditions have already been reported previously\,\cite{rectangular.zeta.}. Also, the leading order of the Casimir energy for the massive scalar field in a rectangular box with $p$ confined sides in $D$ spatial dimensions has already been calculated. This calculation has been done by the zeta function regularization techniques supplemented with reflection formula in\,\cite{wolfram.}. The final answer of this energy per unit length of the waveguide is:
\begin{eqnarray}\label{Leading.Oreder.Cas.}
    E^{(0)}_{\mbox{\footnotesize Cas.}}= \frac{m^2a_1a_2}{8\pi^2}\sum_{j=1}^{\infty}\bigg[\sqrt{\frac{\pi}{m}}\frac{K_{3/2}(2ma_1j)}{a_2(a_1j)^{3/2}}-\frac{K_{2}(2ma_1j)}{(a_1j)^{2}}
    -\sum_{i=1}^{\infty}\frac{K_2(2m\sqrt{(a_1i)^2+(a_2j)^2})}{a_1^2i^2+a_2^2j^2}\bigg]+\{a_1\leftrightarrow a_2\},
\end{eqnarray}
where $a_1$ and $a_2$ are the rectangular cross-sections of the open-ended waveguide and $m$ is the mass of the real scalar field. In this paper\,(Appendix \ref{appendix.zero.order}), we calculated this order of energy for the waveguide via BSS. Two issues in performing this calculation are important. Firstly, the calculation of the leading order Casimir energy is more simple than the first-order ones. Hence, this simplicity provides an opportunity for us to introduce the BSS in a better and also simpler way. Secondly, the result of the leading order Casimir energy in the rectangular waveguide has already been reported in\,\cite{wolfram.}. Therefore, comparing the results of our final answer for the zero-order Casimir energy with the previously reported ones provides a chance in which to examine the merits of BSS. It is noteworthy that our final answer in Appendix\,\ref{appendix.zero.order} agrees with those of Eq.\,\eqref{Leading.Oreder.Cas.} reported in\,\cite{wolfram.}. In the next section, the first-order radiative correction to the Casimir energy for the real massive scalar field in $\phi^4$ theory in a rectangular waveguide is calculated, followed by estimating it in the massless case. We have also discussed appropriate limits of the obtained answers and, in Section\,\ref{sec.conclusion}, summarized all physical aspects of using the applied methods and obtained results.

\section{First Order Radiative Correction to The Casimir Energy}
\label{sec:Cas.Cal.}
In this section, next to the leading order (two-loop quantum correction) of the Casimir energy for the real massive scalar field in $\phi^4$ theory in a rectangular waveguide is calculated. As outlined in the introduction, the main importance of the performed renormalization program in this paper is that the counterterms are systematically extracted from the standard perturbation theory and their expression is position-dependent due to the presence of non-trivial boundary condition. The calculation of such counterterms has been extensively discussed and thus we briefly state the renormalization procedure and conditions\,\cite{posision.dependent.counterterms.works.ours.1,posision.dependent.counterterms.works.ours.2}. Through applying a standard procedure for setting up the renormalized perturbation theory, the Lagrangian of the real massive scalar field with self-interaction term $\phi^4$ after re-scaling the field $\phi=z^{1/2}\phi_r$ becomes:
\begin{equation}\label{lagrangy.}
  \mathcal{L}(x)= \frac{1}{2}\partial_\mu \phi_r(x) \partial^\mu \phi_r(x)-\frac{1}{2}m^2 \phi_r^2(x)-\frac{\lambda}{4!} \phi_{r}^4(x)+\frac{1}{2}\delta_z \partial_\mu \phi_{r}(x) \partial^\mu \phi_{r}(x)-\frac{1}{2}\delta_m \phi_{r}^2(x)-\frac{\delta_\lambda}{4!}\phi_{r}^4(x),
\end{equation}
where $\delta_m$, $\delta_\lambda$, and $\delta_z$ are the counterterms and $x=(t,\mathbf{x})$. Also, the parameters $m$ and $\lambda$ are the physical mass of the field and physical value of coupling constant, respectively. Now, using the usual context of renormalized perturbation theory, the perturbation expansion related to the two-point function, up to the first order of $\lambda$, can be written symbolically as:
\begin{equation}\label{PerturbationEx.}
   \raisebox{-4mm}{\includegraphics[width=1.5cm]{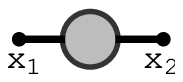}}=\raisebox{-2.7mm}{\includegraphics[width=1.4cm]{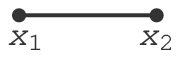}}
   +\raisebox{-2.5mm}{\includegraphics[width=1.2cm]{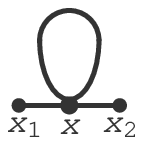}}+\raisebox{-3.5mm}{\includegraphics[width=1.5cm]{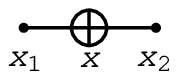}}\hspace{0cm},
\end{equation}
where $\raisebox{-3.mm}{\includegraphics[width=1.5cm]{16.eps}}$ is the counterterm in the above perturbation expansion. After applying the renormalization conditions, the general expression for counterterms becomes:
\begin{equation}\label{Counter-terms.}
   \delta_z=0,\hspace{2cm}\delta_m(x)=\frac{-i}{2}\raisebox{0.2mm}{\includegraphics[width=1cm]{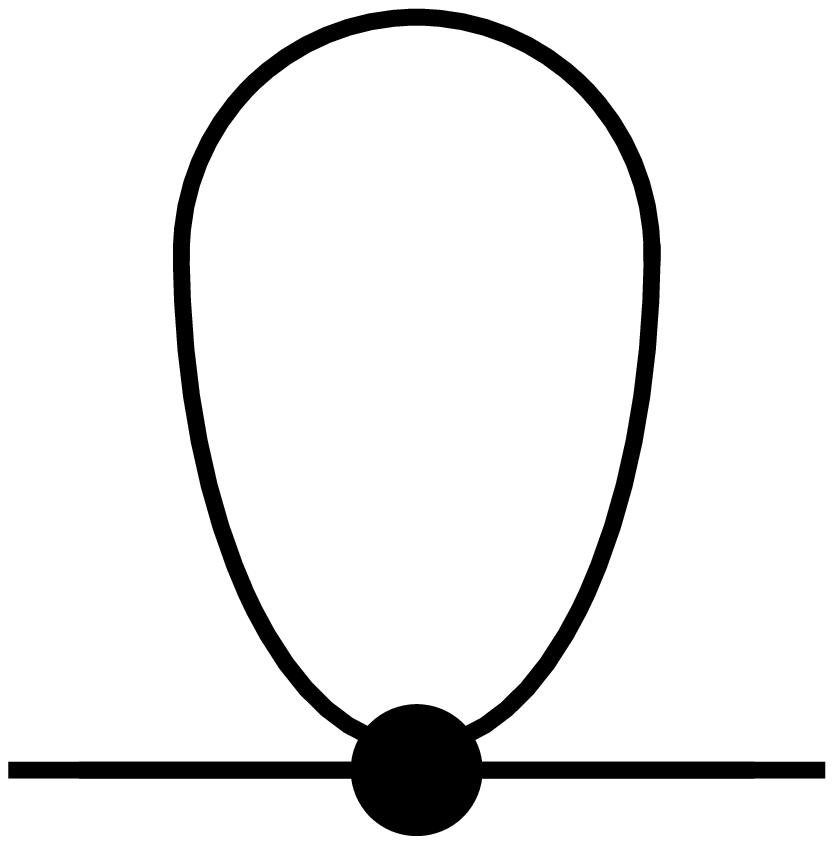}}=\frac{-\lambda}{2}G(x,x),\hspace{2cm}\delta_\lambda=0,
\end{equation}
where $G(x,x)$ is the Green's function. To get the vacuum energy expression per unit length of an open-ended waveguide with a cross-section $a_1\times a_2$\,(region $A1$ in Fig.\,(\ref{fig.2})), we have:
\begin{equation}\label{VacuumEn.damble.}
  E^{(1)}_{A1}=\lim_{L\to\infty}E^{(1)}/L=i\int_{S} dS\bigg(\frac{1}{8} \raisebox{-7mm}{\includegraphics[width=0.5cm]{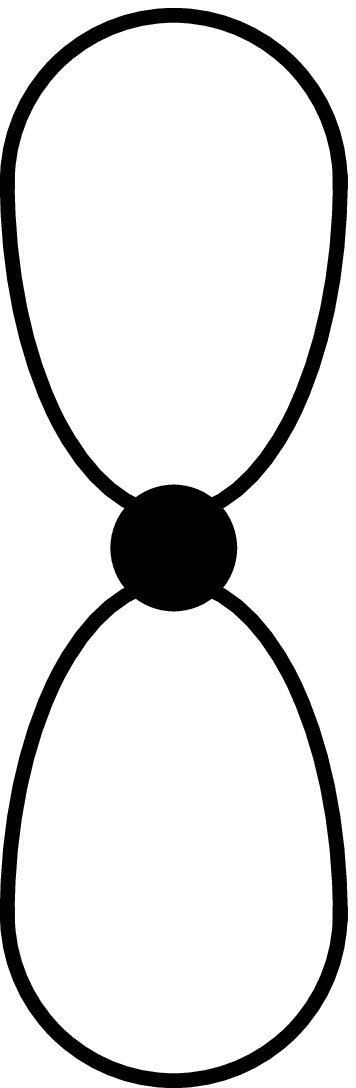}}+\frac{1}{2}\raisebox{-1mm}{\includegraphics[width=0.5cm]{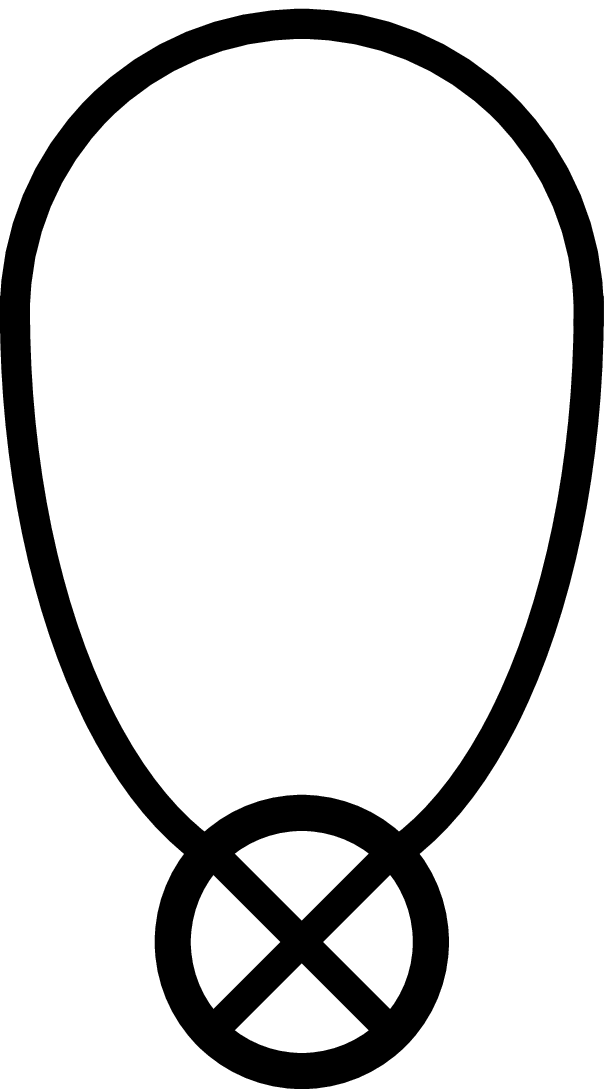}}+...\bigg)=i \int_{S}dS\bigg(\frac{-i\lambda}{8}G^2(x,x)-\frac{-i}{2}\delta_m(x)G(x,x)\bigg).
\end{equation}
This quantity was written up to the first order of coupling constant $\lambda$. The superscript $(1)$ denotes this order for the vacuum energy. The parameter $S$ denotes the area of the cross-section $a_1\times a_2$ of the waveguide. By substituting $\delta_m (x)$ from Eq.\,\eqref{Counter-terms.} for Eq.\,\eqref{VacuumEn.damble.}, the vacuum energy per unit length can be gained. Therefore, we have:
\begin{equation}\label{VacuumEn.firstorder.}
  E^{(1)}_{A1}=\frac{-\lambda}{8} \int_{S} G^2 (x,x)dS.
\end{equation}
According to the lagrangian stated in Eq.\,\eqref{lagrangy.}, the final expression for the Green's function of the massive scalar field with Dirichlet boundary condition in the waveguide with the cross-section $a_1\times a_2$ after Wick rotation can be written as follows:
\begin{eqnarray}\label{Green.function}
  G(x,x')=\frac{4}{a_1a_2}\int\frac{d^2k}{(2\pi)^2}
  \hspace{13cm}\nonumber\\ \times\sum_{n_1,n_2=1}^{\infty}\frac{e^{-\omega(t-t')}e^{-i k_z(z-z')}\sin(\frac{n_1\pi}{a_1}(x+\frac{a_1}{2}))\sin(\frac{n_2\pi}{a_2}(y+\frac{a_2}{2}))\sin(\frac{n_1\pi}{a_1}(x'+\frac{a_1}{2}))
  \sin(\frac{n_2\pi}{a_2}(y'+\frac{a_2}{2}))}{k^2+k_{A1}^2+m^2+i\epsilon},\hspace{0.6cm}
\end{eqnarray}
where $k_{A1}=\sqrt{(\frac{n_1\pi}{a_1})^2+(\frac{n_2\pi}{a_2})^2}$ and $k=(\omega,\mathbf{k}_z)$. Now, by using Eqs.\,\eqref{VacuumEn.firstorder.} and \eqref{Green.function}, the expression for the vacuum energy per unit length of the waveguide for region $A1$ in Fig.\,(\ref{fig.2}) becomes:
\begin{eqnarray}\label{VacuumEn.after.firstintegrations.}
  E^{(1)}_{A1}=\frac{-\lambda}{32\pi^2a_1a_2}\sum_{n_1,n'_1=1}^{\infty}\sum_{n_2,n'_2=1}^{\infty}\bigg[\int_{0}^{\infty}\frac{kdk}{k^2+k_{A1}^2+m^2}\bigg]
  \bigg[\int_{0}^{\infty}\frac{k' dk'}{{k'}^{2}+{k'}_{A1}^2+m^2}\bigg]\nonumber\\
  \times\bigg[1+\frac{1}{2}\delta_{n_1,{n'}_1}+\frac{1}{2}\delta_{n_2,{n'}_2}+\frac{1}{4}\delta_{n_1,{n'}_1}\delta_{n_2,{n'}_2}\bigg],
\end{eqnarray}
where ${k'}_{A1}=\sqrt{(\frac{{n'}_1\pi}{a_1})^2+(\frac{{n'}_2\pi}{a_2})^2}$. This step for calculation of the vacuum energy is repeated for other regions and the obtained expressions would be similar to the one shown in Eq.\,\eqref{VacuumEn.after.firstintegrations.}, except that their parameters might be different. It has to be noted that, according to Eq.\,\eqref{subtraction.vacuums.}, we should calculate the subtraction of vacuum energies for both configurations. However, for the sake of simplicity in reporting the calculation steps, from here on, we only follow the expressions for the original region $A1$ and do not report them for the other regions.
\par
All integrals in Eq.\,\eqref{VacuumEn.after.firstintegrations.} are logarithmically divergent and should be regularized. First, we nondimensionalize them by multiplying factor of $a_1$. Next, using the relation $\sum_{n}\sum_{m}f(x,y)=\sum_{n}\sum_{m}\frac{1}{2}(f(x,y)+f(y,x))$,  the expression for $E^{(1)}_{A1}$ is symmetrized in its double arguments denoted by $a_1$ and $a_2$. Then, the Cutoff Regularization\,(CR) for each integral is performed, by adjusting a separate cutoff for the upper limit of each integral and expanding the result of the integral in the limit in which the cutoffs tend to infinity as follows:
\begin{eqnarray}\label{cal.of.integration.}
  \int_{0}^{\Lambda}\frac{x\,dx}{x^2+a^{2}}=\frac{1}{2}\ln(x^2+a^2)
  \Big|_{0}^{\Lambda}\buildrel {\Lambda  \to\infty} \over
 \longrightarrow \ln\Lambda-\ln a.
\end{eqnarray}
This expansion helps to manifest the explicit form of infinite parts in the integration result. Now, we substitute the expanded form of integration from Eq.\,\eqref{cal.of.integration.} in Eq.\,\eqref{VacuumEn.after.firstintegrations.}. The same procedure for other regions of Fig.\,(\ref{fig.2}) should be conducted analogously. Then, using Eq.\,\eqref{subtraction.vacuums.} and appropriate adjustment of each cutoff, all the infinite parts of Eq.\,\eqref{VacuumEn.after.firstintegrations.} for all regions would cancel each other out due to our BSS\,(for more details see Appendix\,\ref{appendix.CR.technique}). Therefore, we have:
\begin{eqnarray}\label{All.Four.Terms}
   \Delta E^{(1)}_{\mbox{\tiny Vac.}}&=&\frac{-\lambda }{128\pi^2a_1a_2}\nonumber\\
   &\times&\sum_{n_1,n'_1=1}^{\infty}\sum_{n_2,n'_2=1}^{\infty}\ln(\omega_{A1}^2a_1^2)\ln({\omega'}_{A1}^2a_1^2)\bigg[1+\frac{1}{2}\delta_{n_1,{n'}_1}+
   \frac{1}{2}\delta_{n_2,{n'}_2}+\frac{1}{4}\delta_{n_1,{n'}_1}\delta_{n_2,{n'}_2}\bigg]+\{a_1\leftrightarrow a_2\}+...,
\end{eqnarray}
where $\omega_{A1}^2=k_{A1}^2+m^2$ and ${\omega'}_{A1}^2={k'}_{A1}^2+m^2$. As mentioned earlier, only the terms related to region $A1$ in the right side of $\Delta E^{(1)}_{\mbox{\tiny Vac.}}$ were reported. For the sake of simplicity in presenting of the calculation, the reports of expressions related to the other regions in Eq.\,\eqref{All.Four.Terms} were ignored. All summations in Eq.\,\eqref{All.Four.Terms} are still divergent and to regularize their divergences and remove infinities, the CR and BSS are required. For this purpose, the following form of Abel-Plana Summation Formula\,(APSF) is used. This summation formula converts all summation forms in Eq.\,\eqref{All.Four.Terms} into the integral form as follows\,\cite{Generalized.Abel.Plana.Saharian}:
\begin{equation}\label{APSF}
  \sum_{n=1}^{\infty}f(n)=-\frac{1}{2}f(0)+\int_{0}^{\infty}f(z)dz+i\int_{0}^{\infty}\frac{f(it)-f(-it)}{e^{2\pi t}-1}dt,
\end{equation}
where the last term on the right side of Eq.\,\eqref{APSF} is henceforth named as the \emph{Branch-cut} term. Applying APSF and performing all required regularization procedure for Eq.\,\eqref{All.Four.Terms} is a very time-consuming process. For the sake of transparency in the calculation and with respect to the four terms in the bracket of Eq.\,\eqref{All.Four.Terms}, we split the calculations into four parts. So, we briefly follow the calculation for each part in the next sub-sections. Finally, the summary of all these four parts will be discussed.

\subsubsection{The First Term}
\label{sub.section.first.term}
For the first term of Eq.\,\eqref{All.Four.Terms}, after applying the APSF on summations over $n_1$ and ${n'}_1$, we obtain:
\begin{eqnarray}\label{Term1.A1}
  T_1=\frac{-\lambda}{128\pi^2a_1a_2}\sum_{n_1,n'_1=1}^{\infty}\sum_{n_2,n'_2=1}^{\infty}\ln(\omega^2a_1^2)\ln({\omega'}^2a_1^2)\hspace{7cm}\nonumber \\
  = \frac{-\lambda L}{128\pi^2a_1a_2}\sum_{n_2,n'_2=1}^{\infty}\bigg[\frac{-1}{2}\ln((\frac{n_2\pi a_1}{a_2})^2+m^2a_1^2)+\int_{0}^{\infty}dx\ln\Big(x^2\pi^2+(\frac{n_2\pi a_1}{a_2})^2+m^2a_1^2\Big)\hspace{0.7cm}\nonumber\\+\ln\Big(1-e^{-2ma_1\sqrt{(\frac{n_2\pi}{ma_2})^2+1}}\Big)\bigg]\times\bigg[n_2\leftrightarrow {n'}_2\bigg]+\{a_1\leftrightarrow a_2\}.\hspace{2cm}
\end{eqnarray}
Re-applying the APSF converts all remaining summation terms in Eq.\,\eqref{Term1.A1} to the integral form as:
\begin{eqnarray}\label{Term1.A2}
  T_1=\frac{-\lambda}{128\pi^2a_1a_2}\Bigg\{\frac{1}{4}\ln(m^2a_1^2)\underbrace{-\frac{1}{2}\int_{0}^{\infty}dx\ln\Big((\frac{x\pi a_1}{a_2})^2+m^2a_1^2\Big)}_{I_1}+B_1(a_2)\hspace{6cm}\nonumber\\ \underbrace{-\frac{1}{2}\int_{0}^{\infty}dx\ln\Big(x^2\pi^2+m^2a_1^2\Big)}_{I_2}+
  \underbrace{\int_{0}^{\infty}dx\int_{0}^{\infty}dy\ln\Big(x^2\pi^2+(\frac{y\pi a_1}{a_2})^2+m^2a_1^2\Big)}_{I_3}+B_2(a_1,a_2)\hspace{0.5cm}\nonumber\\+B_1(a_1)+
  \underbrace{\frac{ma_2}{\pi}\int_{0}^{\infty}dN\ln\Big(1-e^{-2ma_1\sqrt{N^2+1}}\Big)}_{I_4=B_2(a_2,a_1)}+B_3(a_1,a_2)\Bigg\}^2+\{a_1\leftrightarrow a_2\},\hspace{2.3cm}
\end{eqnarray}
where $B_1(x)$, $B_2(x,y)$, and $B_3(x,y)$ are the Branch-cut term of APSF. All these terms are finite and their expressions are:
\begin{eqnarray}\label{Term1.A3}
  B_1(x)=\frac{-1}{2}\ln(1-e^{-2mx}),\hspace{0.7cm} B_2(x,y)=\frac{-2m^2xy}{\pi}\int_{1}^{\infty}dN\frac{\sqrt{N^2-1}}{e^{2myN}-1},\hspace{0.7cm} B_3(x,y)=-B_1(y)-B_2(x,y).
\end{eqnarray}
The expressions for $B_i$ show that the sum of their contribution in  Eq.\,\eqref{Term1.A2} will be exactly zero. The integral terms indicated by $I_1$, $I_2$, and $I_3$ in the bracket of Eq.\,\eqref{Term1.A2} have a divergent value. To remove their infinities, the CR and BSS should be employed again. Therefore, similar to what occurred for Eq.\,\eqref{VacuumEn.after.firstintegrations.}, the upper limits of each integration are replaced with multiple cutoffs. Then, by calculating integrations, we will have an answer as a function of cutoffs for each integral. In the following, we expand the integration result in the infinite limit of cutoffs in order to manifest the divergent part of each integral. A similar scenario should be conducted for a similar term in the other regions. Now, by adjusting the proper values for cutoffs and using the BSS (Eq.\,\eqref{subtraction.vacuums.}), all divergent parts of these integrals (the term as a function of the cutoff) in Eq.\,\eqref{Term1.A2} would be removed. The remaining finite parts related to each integral term become:
\begin{eqnarray}\label{Term1.A4.Is.}
  I_1\longrightarrow \frac{-ma_2}{2},\hspace{2cm}    I_2\longrightarrow \frac{-ma_1}{2},\hspace{2cm}  I_3\longrightarrow \frac{m^2a_1a_2}{4\pi}(1+\ln4),\hspace{3cm}\nonumber \\
  I_1\times I_1 \longrightarrow \frac{m^2a_2^2}{4\pi^2}(4+\pi^2-2\ln m^2a_1^2),\hspace{2.5cm}
  I_2\times I_2 \longrightarrow \frac{m^2a_1^2}{4\pi^2}(4+\pi^2-2\ln m^2a_1^2),\hspace{1.5cm} \nonumber\\
  I_3\times I_3 \longrightarrow \frac{m^4a_2^2a_1^2}{16\pi^2}(2+2\ln4+\ln^{2}4),\hspace{2.4cm}
  I_1\times\,I_2\longrightarrow  \frac{m^2a_1a_2}{4},\hspace{4.3cm}\nonumber\\
  I_1\times\,I_3\longrightarrow  \frac{-m^3a_1a_2^2}{8\pi}(1+\ln4),\hspace{3.5cm}
  I_2\times\,I_3\longrightarrow  \frac{-m^3a_1^2a_2}{8\pi}(1+\ln4).\hspace{2.5cm}
\end{eqnarray}
It can be shown that the contribution of the finite part of $I_1\times I_2$, $I_1\times I_3$, and $I_2\times I_3$ written in Eq.\,\eqref{Term1.A4.Is.} will be also removed due to our BSS. By substituting the remaining terms for Eq.\,\eqref{Term1.A2}, the final expression $T_1$ for region $A1$ is obtained. It has to be noted that, similar to any terms remained in $T_1$, we will obtain for the other regions. However, the reporting of them were ignored here, for the sake of simplicity.

\subsubsection{The Second Term}
For the second term of Eq.\,\eqref{All.Four.Terms}, after applying the APSF on both summations over $n_1$ and ${n'}_1$, we obtain:
\begin{eqnarray}\label{Term2.A1.}
  T_2=\frac{-\lambda}{128\pi^2a_1a_2}\frac{1}{2}\sum_{n_2=1}^{\infty}\sum_{n_1,n'_1=1}^{\infty}\ln(\omega^2a_1^2)\ln({\omega'}^2a_1^2)\hspace{8cm}\nonumber \\= \frac{-\lambda L}{128\pi^2a_1a_2}\frac{1}{2}\sum_{n_2=1}^{\infty}\bigg[\frac{-1}{2}\ln\Big((\frac{n_2\pi a_1}{a_2})^2+m^2a_1^2\Big)+\underbrace{\int_{0}^{\infty}dx\ln\Big(x^2\pi^2+(\frac{n_2\pi a_1}{a_2})^2+m^2a_1^2\Big)}_{I_5}\hspace{1.4cm}\nonumber\\+\ln\Big(1-e^{-2ma_1\sqrt{(\frac{n_2\pi}{ma_2})^2+1}}\Big)\bigg]^2+\{a_1\leftrightarrow a_2\}.\hspace{4.9cm}
\end{eqnarray}
The term $I_5$ for any finite values of $n_2$ is divergent. To remove its infinity, similar to what occurred for the integral terms in Eq.\,\eqref{Term1.A2}, we repeat all steps of CR and BSS. After employing that scenario, the following finite parts for $I_5$ will be obtained:
\begin{eqnarray}\label{Term2.A2.Is.1}
  I_5\longrightarrow ma_1\sqrt{(\frac{n_2\pi}{ma_2})^2+1},\hspace{2cm}    I_5\times I_5\longrightarrow \frac{m^2a_1^2}{\pi^2}\Big((\frac{n_2\pi}{ma_2})^2+1\Big)\Big(4+\pi^2-2\ln(m^2a_1^2)\Big).
\end{eqnarray}
\vspace{-0.2cm}
Now by re-applying the APSF on the remaining summation of Eq.\,\eqref{Term2.A1.}, we have:
\begin{eqnarray}\label{Term2.A3.}
  T_2=\frac{-\lambda}{128\pi^2a_1a_2}\frac{1}{2}\Bigg[\frac{-1}{8}\ln^2(m^2a_1^2)+\underbrace{\frac{1}{4}\int_{0}^{\infty}dx\ln^2\Big((\frac{x\pi a_1}{a_2})^2+m^2a_1^2\Big)}_{I_6}+B_4(a_2)\hspace{5cm}\nonumber\\
  +\frac{1}{2}ma_1\ln(m^2a_1^2)\underbrace{-ma_1\int_{0}^{\infty}dx\sqrt{(\frac{x\pi}{ma_2})^2+1}\ln((\frac{x\pi a_1}{a_2})^2+m^2a_1^2)}_{I_7}+B_5(a_1,a_2)\hspace{2cm}\nonumber\\
  -\ln(m^2a_1^2)B_1(a_1)\underbrace{-\int_{0}^{\infty}dx\ln\Big(1-e^{-2ma_1\sqrt{(\frac{x\pi}{ma_2})^2+1}}\Big)\ln\Big((\frac{x\pi a_1}{a_2})^2+m^2a_1^2\Big)}_{-2\ln ma_1 B_2(a_2,a_1)+I_8(a_2,a_1)}+B_6(a_1,a_2)\hspace{0.3cm}\nonumber\\
  +2ma_1B_1(a_1)+\underbrace{\frac{2m^2a_1a_2}{\pi}\int_{0}^{\infty}dN\ln\Big(1-e^{-2ma_1\sqrt{N^2+1}}\Big) \sqrt{N^2+1}}_{I_9(a_1,a_2)}+B_7(a_1,a_2)\hspace{1.8cm}\nonumber\\
  -2B_1^2(a_1)+\underbrace{\frac{ma_2}{\pi}\int_{0}^{\infty}dN\ln^2\Big(1-e^{-2ma_1\sqrt{N^2+1}}\Big)}_{I_{10}(a_1,a_2)}+B_8(a_1,a_2)\hspace{4.4cm}\nonumber\\
  +\frac{m^2a_1^2}{\pi^2}(4+\pi^2-2\ln(m^2a_1^2))\Big[\frac{-1}{2}+\underbrace{\frac{ma_2}{\pi}\int_{0}^{\infty}dN(N^2+1)}_{I_{11}(a_2)}\Big]+B_9  \Bigg]+\{a_1\leftrightarrow a_2\},\hspace{1.5cm}
\end{eqnarray}
where $\dd I_8(x,y)=\frac{-mx}{\pi}\int_{0}^{\infty}dN\ln(N^2+1)\ln(1-e^{-2my\sqrt{N^2+1}})$. All terms denoted by $B_i$ in Eq.\,\eqref{Term2.A3.} are the Branch-cut terms of APSF and their values are:
\begin{eqnarray}\label{Term2.A4.Bs}
   B_4(x)&=&-\ln(m^2a_1^2)B_1(x)-mx\int_{1}^{\infty}dN\frac{\ln(N^2-1)}{e^{2mxN}-1},\nonumber\\ B_5(x,y)&=&-\ln(m^2a_1^2)B_2(x,y)+\frac{2m^2xy}{\pi}\int_{1}^{\infty}dN\frac{\sqrt{N^2-1}\ln(N^2-1)}{e^{2myN}-1},\nonumber\\
   B_6(x,y)&=&-B_4(y)-B_5(x,y)+my\int_{1}^{\infty}dN\frac{\ln(4\sin^2(mx\sqrt{N^2-1}))}{e^{2myN}-1},\nonumber\\
   B_7(x,y)&=&\frac{-2m^2xy}{\pi}\int_{1}^{\infty}\frac{\sqrt{N^2-1}\ln(4\sin^2(mx\sqrt{N^2-1}))}{e^{2myN}-1},\nonumber\\
   B_8(x,y)&=&-B_7(x,y)-B_6(x,y)-B_5(x,y)-B_4(y),\hspace{2cm}     B_9=0.
\end{eqnarray}
According to the expressions for Branch-cut terms in  Eq.\,\eqref{Term2.A4.Bs}, it can be easily shown that their summation is exactly zero. In fact, they will not leave any contribution in $T_2$. The integrals $I_6$, $I_7$, and $I_{11}(x)$ have a divergent value. Thus, to regularize these terms and remove their infinities, the CR supplemented by BSS should be used. The details of this scenario of regularization were discussed in the previous lines. Therefore, we only reported the remaining finite parts of $I_6$, $I_7$, and $I_{11}(x)$ as follows:
\begin{eqnarray}\label{Term2.A5.Is.}
    \hspace{-1cm} I_6\longrightarrow \frac{ma_2}{2}(\ln (m^2a_1^2)+2\ln2-2),\hspace{1cm}     I_7\longrightarrow\frac{-m^2a_1a_2}{4\pi}\ln (m^2a_1^2)(1+\ln4),\hspace{1cm}
     I_{11}(a_2)\longrightarrow 0.
\end{eqnarray}
The other integrals in Eq.\,\eqref{Term2.A3.} are convergent and their contributions ultimately remain in the Casimir energy.

\subsubsection{The Third Term}
For the third term of Eq.\,\eqref{All.Four.Terms}, after applying the APSF on both summations over $n_2$ and ${n'}_2$, we obtain:
\begin{eqnarray}\label{Term3.A1.}
  T_3=\frac{-\lambda}{128\pi^2a_1a_2}\frac{1}{2}\sum_{n_1=1}^{\infty}\sum_{n_2,n'_2=1}^{\infty}\ln(\omega^2a_1^2)\ln({\omega'}^2a_1^2)\hspace{6cm}\nonumber \\
  = \frac{-\lambda}{128\pi^2a_1a_2}\frac{1}{2}\sum_{n_1=1}^{\infty}\bigg[\frac{-1}{2}\ln\Big((n_1\pi)^2+m^2a_1^2\Big)+
  \underbrace{\int_{0}^{\infty}dx\ln\Big(n_1^2\pi^2+(\frac{x\pi a_1}{a_2})^2+m^2a_1^2\Big)}_{I_{12}}\nonumber\\+\ln\Big(1-e^{-2ma_2\sqrt{(\frac{n_{1}\pi}{ma_1})^2+1}}\Big)\bigg]^2+\{a_1\leftrightarrow a_2\}.\hspace{3cm}
\end{eqnarray}
The term $I_{12}$ for any finite values of $n_1$ is divergent. Using the CR and BSS, we can remove its infinity and thus the remaining finite expressions related to $I_{12}$ become:
\begin{eqnarray}\label{Term3.A2.Is.1}
  I_{12}\longrightarrow ma_2\sqrt{(\frac{n_1\pi}{ma_1})^2+1},\hspace{2cm}    I_{12}\times I_{12}\longrightarrow \frac{m^2a_2^2}{\pi^2}\Big((\frac{n_1\pi}{ma_1})^2+1\Big)\Big(4+\pi^2-2\ln(m^2a_1^2)\Big).
\end{eqnarray}
Now, by substituting Eq.\,\eqref{Term3.A2.Is.1} in appropriate places of Eq.\,\eqref{Term3.A1.} and applying APSF again on the remaining summation of Eq.\,\eqref{Term3.A1.} we have:
\begin{eqnarray}\label{Term3.A3.}
  T_{3}=\frac{-\lambda
  }{128\pi^2a_1a_2}\frac{1}{2}\Bigg[\frac{-1}{8}\ln^2(m^2a_1^2)+\underbrace{\frac{1}{4}\int_{0}^{\infty}dx\ln^2\Big((x\pi)^2+m^2a_1^2\Big)}_{I_{13}}+B_4(a_1)\hspace{5cm}\nonumber\\
  +\frac{1}{2}ma_2\ln(m^2a_1^2)\underbrace{-ma_2\int_{0}^{\infty}dx\sqrt{(\frac{x\pi}{ma_1})^2+1}\ln((x\pi)^2+m^2a_1^2)}_{I_{14}}+B_5(a_2,a_1)\hspace{2cm}\nonumber\\
  -\ln(m^2a_1^2)B_1(a_2)\underbrace{-\int_{0}^{\infty}dx\ln\Big(1-e^{-2ma_2\sqrt{(\frac{x\pi}{ma_1})^2+1}}\Big)
  \ln\Big((x\pi)^2+m^2a_1^2\Big)}_{-2\ln ma_1 B_2(a_1,a_2)+I_8(a_1,a_2)}+B_6(a_2,a_1)\hspace{0.3cm}\nonumber\\
  +2ma_2B_1(a_2)+\underbrace{\frac{2m^2a_1a_2}{\pi}\int_{0}^{\infty}dN\ln\Big(1-e^{-2ma_2\sqrt{N^2+1}}\Big)
  \sqrt{N^2+1}}_{I_9(a_2,a_1)}+B_7(a_2,a_1)\hspace{1.4cm}\nonumber\\
  -2B_1^2(a_2)+\underbrace{\frac{ma_1}{\pi}\int_{0}^{\infty}dN\ln^2\Big(1-e^{-2ma_2\sqrt{N^2+1}}\Big)}_{I_{10}(a_2,a_1)}+B_8(a_2,a_1)\hspace{4cm}\nonumber\\
  +\frac{m^2a_2^2}{\pi^2}(4+\pi^2-2\ln(m^2a_1^2))\Big[\frac{-1}{2}+\underbrace{\frac{ma_1}{\pi}\int_{0}^{\infty}(N^2+1)dN}_{I_{11}(a_1)}\Big]+B_9  \Bigg]+\{a_1\leftrightarrow a_2\},\hspace{1.2cm}
\end{eqnarray}
The sum of all Branch-cut terms denote by $B_i$ in above expression, analogous to what occurred in the same terms in $T_2$, is exactly zero. In Eq.\,\eqref{Term3.A3.} except for integrals $I_{13}$, $I_{14}$, and $I_{11}(x)$, the other integrations have a finite value. To remove the infinite part of these three terms, the CR and BSS should be used. Afterward, the remaining finite part of each term becomes:
\begin{eqnarray}\label{Term3.A4.Is.}
     I_{13}\longrightarrow \frac{ma_1}{2}(\ln (m^2a_1^2)+2\ln2-2),\hspace{1cm}     I_{14}\longrightarrow\frac{-m^2a_1a_2}{4\pi}\ln (m^2a_1^2)(1+\ln4)\hspace{1cm}
     I_{11}(x)\longrightarrow 0.
\end{eqnarray}

\subsubsection{The Fourth Term}
Applying the APSF on both summations over $n_1$ and $n_2$ converts the last term of Eq.\eqref{All.Four.Terms} to:
\begin{eqnarray}\label{Term4.A1.}
  T_4=\frac{-\lambda}{128\pi^2a_1a_2}\frac{1}{4}\sum_{n_1=1}^{\infty}\sum_{n_2=1}^{\infty}\ln^2(\omega^2a_1^2)\nonumber\hspace{10cm} \\
  =\frac{-\lambda}{128\pi^2a_1a_2}\frac{1}{4}\bigg[\frac{1}{4}\ln^2(m^2a_1^2) \underbrace{-\frac{1}{2}\int_{0}^{\infty}dx\ln^2\Big((\frac{x\pi a_1}{a_2})^2+m^2a_1^2\Big)}_{-2I_6}-2B_4(a_2)\hspace{4.2cm}\nonumber\\
  \underbrace{-\frac{1}{2}\int_{0}^{\infty}dx\ln^2(x^2\pi^2+m^2a_1^2)}_{-2I_{13}}+\underbrace{\int_{0}^{\infty}dx\int_{0}^{\infty}dy\ln^2 (x^2\pi^2+(\frac{y\pi a_1}{a_2})^2+m^2a_1^2)}_{I_{15}}+B_{10}(a_1,a_2)\hspace{0cm}\nonumber\\
  -2B_4(a_1)+B_{10}(a_2,a_1)-4ma_2\int_{0}^{1}\frac{\ln\Big(1-e^{-2ma_1\sqrt{1-N^2}}\Big)}{e^{2ma_2N}-1}
  \bigg]+\{a_1\leftrightarrow a_2\},\hspace{2.2cm}
\end{eqnarray}
where $B_{10}(x,y)=2(\ln4-2)B_2(x,y)-2B_5(x,y)$. The integral $I_{15}$ is divergent and by using CR and BSS the remaining finite part of this term becomes: $\frac{m^2a_1a_2}{12\pi}(-6+7\pi^2+12\ln^22+12(\ln4+1)\ln ma_1)$. The last integral in Eq.\,\eqref{Term4.A1.} is also divergent due to its lower limit. To reveal the type of this divergence, we first replaced the lower limit of integral by a regulator $\epsilon$ and then we expanded the integral result in the limit $\epsilon\to0^{+}$:
\begin{eqnarray}\label{Term4.A3.LastB}
   -4ma_2\int_{\epsilon}^{1}dN\frac{\ln\Big(1-e^{-2ma_1\sqrt{1-N^2}}\Big)}{e^{2ma_2N}-1}\hspace{0.2cm}\buildrel {\epsilon\to 0^{+}}\over \longrightarrow \hspace{0.2cm}2\ln[1-e^{-2ma_1}]\ln\epsilon+\mathcal{O}(\epsilon)
 \end{eqnarray}
This expansion in the limit of $\epsilon\to 0^{+}$ helps to manifest the infinite part of the integral. Obtaining a closed form for the integral result is a highly cumbersome task. Hence, we compute it numerically and then, to get a finite answer from the integral, we subtract\,(or exclude) the contribution of the divergent part\,(the first term in the right hand side of Eq.\,\eqref{Term4.A3.LastB}) from the obtained result of the integral. A similar computation was also conducted for the same term in the other regions. In fact, the parameter $\epsilon$ plays a regulatory role in removing the divergent part of the integral, which should be considered in the limit $\epsilon\to0^{+}$.
\par
In the above sub-sections, four parts of Eq.\,\eqref{All.Four.Terms} were discussed and all infinite parts of them were removed by their counterparts in the other regions. It can be shown that for each region by summing up all remaining terms of four expressions $T_1$, $T_2$, $T_3$, and $T_4$ according to Eq.\,\eqref{All.Four.Terms}, many cancellations also occur internally. All remaining terms, at this step, are convergent for any finite values of $a_1$, $a_2$, $R$, and $m\neq0$. Finally, the limits $R/b\to\infty$ and $b/a\to\infty$ should be calculated according to Eq.\,\eqref{BSS.CAS.Def.}. This limit renders that these remaining terms from all regions except for $A1$ tend to be zero\,(for more details see Appendix\,\ref{appendix.outer.region}). Eventually, the first order radiative correction to the Casimir energy per unit length for massive scalar field confined in an infinite rectangular waveguide with a cross-section $a_1\times a_2$ becomes:
\begin{eqnarray}\label{Final.RC.Massive.}
     E_{\mbox{\small{Cas.}}}^{(1)}&=&\frac{-\lambda}{128\pi^2 a_1 a_2}\bigg\{\Big[B_2(a_2,a_1)+\ln(ma_1)-m(a_1+a_2)+\frac{m^2a_1a_2}{2\pi}(1+\ln4)
     -\underbrace{\ln(1-e^{-2ma_1})}_{\mathcal{K}_1}\Big]B_2(a_2,a_1)
     \nonumber\\
     &+&B_1(a_1)\Big[\frac{m^2a_1a_2}{4\pi}(1+\ln4)+\ln (ma_1)-\frac{ma_2}{2}\Big]+B_1(a_2)\Big[ma_2-B_1(a_2)\Big] +\underbrace{ma_1\int_{1}^{\infty}d\eta\frac{\ln(\eta^2-1)}{e^{2ma_1\eta}-1}}_{\mathcal{K}_2}
     \nonumber\\
     &+&I_8(a_1,a_2)+I_9(a_1,a_2)+I_{10}(a_1,a_2)+B_2(a_1,a_2) - \frac{2m^2a_1a_2}{\pi}\int_{1}^{\infty}d\eta\frac{\sqrt{\eta^2-1}\ln(\eta^2-1)}{e^{2ma_2\eta}-1}
     \nonumber\\
     &-&ma_2\int_{\epsilon}^{1}dN\frac{\ln\Big(1-e^{-2ma_1\sqrt{1-N^2}}\Big)}{e^{2ma_2N}-1}-\frac{1}{2}\ln[1-e^{-2ma_1}]\ln\epsilon\bigg\}
     +\{a_1\leftrightarrow a_2\},\hspace{3.6cm}
\end{eqnarray}
It is of note that the above result for the Casimir energy per unit length of the waveguide is finite for any values of mass $m\neq0$ and its computation should be partly conducted numerically.
\par
The direct calculation for the massless case is highly cumbersome and it is plagued with multiple kinds of divergences and ambiguities. Hence, we start with Eq.\,\eqref{Final.RC.Massive.}, when the limit $R/b\to \infty$ and $b/a\to \infty$ were not been computed. Then, by putting the parameter $m$ as the regulator, we expand each term in the limit $m\to0^{+}$. In this limit, two types of terms denoted by $\mathcal{K}_1$ and $\mathcal{K}_2$ in Eq.\,\eqref{Final.RC.Massive.} are divergent. Their expanded form in the limit $m\to0^{+}$ becomes:
\begin{equation}\label{Basteh.LN.}
    \mathcal{K}_1=\ln(1-e^{-2mx})\approx\ln(2mx)-mx+\mathcal{O}(m^2),\hspace{1.2cm}  \mathcal{K}_2=\frac{ma_1}{8}\int_{1}^{\infty}d\eta\frac{\ln(\eta^2-1)}{e^{2ma_1\eta}-1}\approx\frac{1}{16}\Big[\ln^2(2ma_1)+0.63492\Big].
\end{equation}
\begin{figure}[th] \hspace{0cm}\includegraphics[width=9cm]{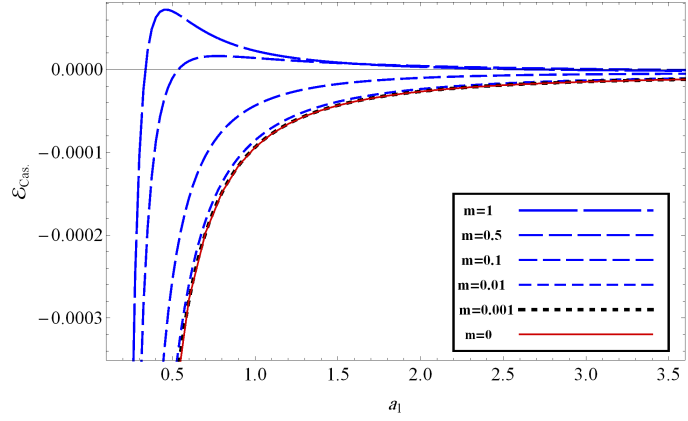}\caption{\label{fig.MasslessCheck} \small
  The plot of the first-order radiative correction to the Casimir energy density for the massive scalar field in a rectangular waveguide with a cross-section $a_1\times a_2$; in this plot, the Casimir energy value in the waveguide per unit volume as a function of the side $a_1$ for sequence values of mass $m=\{1,0.5,0.1,0.01,0.001,0\}$ were displayed\,($a_2=1$ and $\lambda=0.1$). The plot also shows that the Casimir energy for the massive cases converges to the massless case when $m$ decreases and there is an insignificant difference between the figures of the massive cases for $m < 0.01$ and the massless one.}
\end{figure}

\begin{figure}[th] \hspace{0cm}\includegraphics[width=9cm]{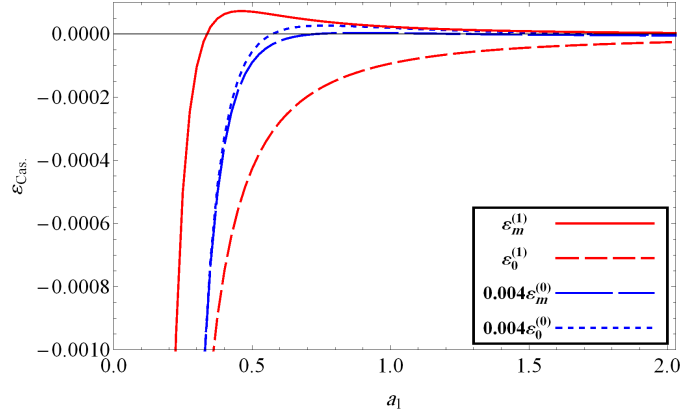}\caption{\label{fig.4plot} \small
 The leading term for the Casimir energy per unit volume (multiplied by a factor of $1/250$) and its first-order radiative correction in the rectangular waveguide with a cross-section $a_1\times a_2$ plotted as a function of the side $a_1$ for a massive ($m = 1$) and massless scalar fields for $\lambda = 0.1$.}
\end{figure}
Substituting above expansions in Eq.\,\eqref{Final.RC.Massive.} manifests the divergent parts of that equation in the limit $m\to0^{+}$. Now, by adjusting the parameters $b_1$, $b_2$, $R$, and $m$ appropriately, we are able to remove all infinities via the BSS and CR technique. The final expression for the first-order radiative correction to the Casimir energy per unit length of the waveguide for the massless scalar field is derived numerically as:
\begin{equation}\label{Final.RC.Massless.Cas.Energy.}
    E^{(1)}_{\mbox{\small{Cas.}}}\approx \frac{-\lambda}{32\pi^2 a_1 a_2}\Biggl[\Big(\frac{\pi a_2}{24a_1}\Big)^2+\Big(\frac{\pi a_1}{24a_2}\Big)^2+\frac{\pi}{12}\Biggr].
\end{equation}

\par
Fig.\,(\ref{fig.MasslessCheck}) presents the first-order correction to the Casimir energy density for different values of mass $m$. This figure exhibits the Casimir energy value for the massive scalar field converges into the massless case when the limit $m\to0^{+}$. Fig.\,(\ref{fig.4plot}) shows the results for leading and next to the leading order of the Casimir energy density for massive and massless scalar fields in a rectangular waveguide. As can be seen, the zero-order Casimir energy is $250$ times larger than its first-order value for $\lambda=0.1$.
\par
The other extreme limit of the waveguide is when one side of the waveguide approaches to infinity, for example $a_2$. As expected, in this limit, the Casimir energy value should convert to that of for two parallel plates. By calculating the limit $a_2\to\infty$ in Eq.\,\eqref{Final.RC.Massive.}, the following form of the Casimir energy density is obtained:
\begin{eqnarray}\label{two.plate.limit.massive}
    \mathcal{E}_{\mbox{\small{Cas.}}}^{(1)}=E/(L a_2)=\frac{-\lambda}{128\pi^3}\sum_{j=1}^{\infty}\frac{K_1(2ma_1j)}{j} \bigg[\frac{m}{\pi a_1}\sum_{j'=1}^{\infty}\frac{K_1(2ma_1j')}{j'}+\frac{m}{a_1}-\frac{m^2}{2\pi}(1+\ln4)\bigg]
\end{eqnarray}
This result for the Casimir energy density is exactly compatible with those reported in previous works\,\cite{posision.dependent.counterterms.works.ours.2}. Similarly, for the massless case in the limit $a_2\to\infty$, we have:
\begin{eqnarray}\label{two.plate.massless}
\mathcal{E}_{\mbox{\small{Cas.}}}^{(1)}=E/(L a_2)=\frac{-\lambda}{18432a_1^3}
\end{eqnarray}
This result is also consistent with a previously reported result\,\cite{posision.dependent.counterterms.works.ours.2} for two parallel plates. These extreme limits can effectively explain the correctness of our obtained result.

\section{Conclusion}
\label{sec.conclusion}
In this paper, we computed the zero- and first-order radiative corrections to the Casimir energy of massive scalar field with the Dirichlet boundary condition in an open-ended rectangular waveguide. In our renormalization program, the counterterm was position-dependent and thus its deduction was conducted by a systematic perturbation theory. This counterterm allows all influences from boundary conditions be imported in the renormalization program. Thus, maintaining that enables us to renormalize all bare parameters of the theory. The other important point in this calculation procedure is applying the Box Subtraction Scheme as a regularization technique. In this regularization procedure, two similar configurations were defined and their vacuum energies were subtracted from each other. This subtraction procedure supplemented by cutoff regularization provides a situation that all infinities cancel out each other without applying any analytic continuation. All the obtained results are consistent with the previously reported results and are also in agreement with physical principles.

\appendix
\section{Calculation of Leading Order Casimir Energy}
\label{appendix.zero.order}
The vacuum energy of free massive scalar field confined with Dirichlet boundary condition in an open-ended rectangular waveguide with a cross-section $a_1\times a_2$ per unit length is expressed as:
\begin{eqnarray}\label{vacuum.energy.zero.order}
     E^{(0)}_{A1}=\frac{1}{2}\int\frac{dk_{z}}{2\pi}\sum_{n_1,n_2=1}^{\infty}\omega(a_1,a_2),
\end{eqnarray}
where $\omega^2(a_1,a_2)=(\frac{n_1\pi}{a_1})^2+(\frac{n_2\pi}{a_2})^2+k_z^2+m^2$ is the wave number. At the first step, using the relation $\sum_{n}\sum_{m}f(x,y)=\sum_{n}\sum_{m}\frac{1}{2}(f(x,y)+f(y,x))$, we symmetrize the expression $E^{(0)}_{A1}$ in its double arguments denoted by $a_1$ and $a_2$. Then, to subtract the vacuum energies of regions displayed in Fig.\,(\ref{fig.2}), starting with Eq.\,\eqref{subtraction.vacuums.}, we obtain:
\begin{eqnarray}\label{subtraction.vacuum.zero.order}
      \Delta E^{(0)}_{\mbox{\tiny Vac.}}=E^{(0)}_{A}- E^{(0)}_{B}&=&\Bigg\{\frac{1}{4}\int\frac{dk_{z}}{2\pi}\bigg[\sum_{n_1,n_2=1}^{\infty}\omega(a_1,a_2)
      +2\sum_{n_1,n_2=1}^{\infty}\omega\big(\frac{R-a_1}{2},\frac{R+a_2}{2}\big)\nonumber\\
      &+&2\sum_{n_1,n_2=1}^{\infty}\omega\big(\frac{R-a_2}{2},\frac{R+a_1}{2}\big)\bigg]
      +\{a_1\leftrightarrow a_2\}\Bigg\}-\{a_1\rightarrow b_1,a_2\rightarrow b_2\}.
\end{eqnarray}
High-frequency render all summations in Eq.\,\eqref{subtraction.vacuum.zero.order} formally divergent. To regularize them, we convert all summation forms into the integral form using the APSF given in Eq.\,\eqref{APSF}. Therefore, we have:
\begin{eqnarray}\label{After.first.APSF.Zero.order}
     E^{(0)}_{A}- E^{(0)}_{B}&=&\Bigg\{\bigg[\frac{-1}{8}\sum_{n_2=1}^{\infty}\int\frac{dk_z}{2\pi}\sqrt{\big(\frac{n_2\pi}{a_2}\big)^2+k_z^2+m^2}
     +\frac{1}{4}\sum_{n_2=1}^{\infty}\int\frac{dk_z}{2\pi}\int_{0}^{\infty}\sqrt{\big(\frac{x\pi}{a_1}\big)^2+\big(\frac{n_2\pi}{a_2}\big)^2+k_z^2+m^2}dx
     \nonumber\\ &+&\mathcal{B}_3(a_1,a_2)\bigg]+2\Big[a_1\rightarrow\frac{R-a_1}{2},a_2\rightarrow\frac{R+a_2}{2}\Big]
     +2\Big[a_1\rightarrow\frac{R-a_2}{2},a_2\rightarrow\frac{R+a_1}{2}\Big]+\{a_1\leftrightarrow a_2\}\Bigg\}\nonumber\\
     &-&\{a_1\rightarrow b_1,a_2\rightarrow b_2\},
\end{eqnarray}
Two terms and their counterparts related to the other regions in the right side of above equation are still divergent. To regularize them, we should apply the APSF again. Therefore, we obtain:
\begin{eqnarray}\label{After.Second.APSF.Zero.order}
     E^{(0)}_{A}- E^{(0)}_{B}&=&\Bigg\{\bigg[\frac{1}{16}\int\frac{dk_z}{2\pi}\sqrt{k_z^2+m^2}
     -\frac{a_2}{8\pi}\underbrace{\int\frac{dk_z}{2\pi}\int_{0}^{\infty}d\xi\sqrt{\xi^2+k_z^2+m^2}}_{\mathcal{J}_1=\frac{1}{2}\int_{0}^{\infty}r\sqrt{r^2+m^2}dr}
     +\mathcal{B}_1(a_2)\nonumber\\
     &-&\frac{a_1}{8\pi}\int\frac{dk_z}{2\pi}\int_{0}^{\infty}d\xi\sqrt{\xi^2+k_z^2+m^2}
     \underbrace{+\frac{1}{4}\int\frac{dk_z}{2\pi}\int_{0}^{\infty}dx\int_{0}^{\infty}dy\sqrt{\big(\frac{x\pi}{a_1}\big)^2+\big(\frac{y\pi}{a_2}\big)^2+k_z^2+m^2}}_{\mathcal{J}_2(a_1,a_2)}
     \nonumber\\ &+&\mathcal{B}_2(a_1,a_2)+\mathcal{B}_3(a_1,a_2)\bigg]+2\Big[a_1\rightarrow\frac{R-a_1}{2},a_2\rightarrow\frac{R+a_2}{2}\Big]\nonumber\\
     &+&2\Big[a_1\rightarrow\frac{R-a_2}{2},a_2\rightarrow\frac{R+a_1}{2}\Big]+\{a_1\leftrightarrow a_2\}\Bigg\}
     -\{a_1\rightarrow b_1,a_2\rightarrow b_2\},\nonumber\\
\end{eqnarray}
where $\mathcal{B}_1(x)$, $\mathcal{B}_2(x,y)$, and $\mathcal{B}_3(x,y)$ are the Branch-cut terms of APSF, which are obtained as follows:
\begin{eqnarray}\label{Branch-cuts.Zero.order}
     \mathcal{B}_1(x)&=&\frac{x}{4\pi^2}\int_{m}^{\infty}d\eta\int_{0}^{\sqrt{\eta^2-m^2}}dk_z\frac{\sqrt{\eta^2-k_z^2-m^2}}{e^{2x\eta}-1}
     =\frac{1}{64\pi x^2}\sum_{j=1}^{\infty}\frac{(1+2mxj)e^{-2mxj}}{j^3},\nonumber\\
     \mathcal{B}_2(x,y)&=&\frac{-xy}{4\pi^2}\int_{m}^{\infty}d\eta\int_{0}^{\sqrt{\eta^2-m^2}}dr\frac{r\sqrt{\eta^2-r^2-m^2}}{e^{2y\eta}-1}
     =\frac{-xm^2}{16\pi^2y}\sum_{j=1}^{\infty}\frac{K_2(2myj)}{j^2},\nonumber\\
     \mathcal{B}_3(x,y)&=&\frac{-x}{2\pi^2}\sum_{n_2=1}^{\infty}\int_{S_{n_2}(y)}^{\infty}d\eta\int_{0}^{\sqrt{\eta^2-S_{n_2}^2(y)}}dk_z
     \frac{\sqrt{\eta^2-k_z^2-S_{n_2}^2(y)}}{e^{2x\eta}-1}
     =\frac{-1}{32\pi x^2}\sum_{j=1}^{\infty}\frac{\big(1+2xjS_{n_2}(y)\big)e^{-2xjS_{n_2}(y)}}{j^3},\nonumber\\
\end{eqnarray}
where $S_{n_2}(y)=\sqrt{\big(\frac{n_2\pi}{y}\big)^2+m^2}$, and $K_2(\alpha)$ is the modified Bessel function. The first term in the right side of Eq.\,\eqref{After.Second.APSF.Zero.order} is divergent. The contribution of this term and its counterparts in the other regions will be automatically removed via subtraction process defined in BSS. The term $\mathcal{J}_1$ expressed in Eq.\,\eqref{After.Second.APSF.Zero.order} is also divergent. As shown in the following, when the subtraction process is performed, all infinite terms cancel each other due to the $\mathcal{J}_1$ originated from outer regions of two configurations $A'$ and $B'$:
\begin{eqnarray}\label{J1(x)-f2}
       \left[2\Big(\frac{R-a_1}{2}+\frac{R+a_2}{2}\Big)+2\Big(\frac{R-a_2}{2}+\frac{R+a_1}{2}\Big)-\{a_1\to b_1,a_2\to b_2\}\right]\left(\frac{-1}{16\pi}\int_{0}^{\infty}r\sqrt{r^2+m^2}dr\right)=0.
\end{eqnarray}
Divergences originated from $\mathcal{J}_1$ for inner regions of configuration $A'$ and $B'$\,(region $A1$ and $B1$) are still left. Therefore, to regularize and remove them, we use the Cutoff Regularization\,(CR) as a supplementary regularization technique. Hence, we replace the upper limit of integrals in $\mathcal{J}_1$ by separate cutoffs $\Lambda_{A1}$ and $\Lambda_{B1}$ for regions $A1$ and $B1$, respectively. So, we have:
\begin{eqnarray}\label{J1(x)-f1}
    &&\frac{-(a_1+a_2)}{16\pi}\int_{0}^{\Lambda_{A1}}r\sqrt{r^2+m^2}dr
    -\frac{-(b_1+b_2)}{16\pi}\int_{0}^{\Lambda_{B1}}r\sqrt{r^2+m^2}dr\nonumber\\
    &&\hspace{2cm}= \frac{-(a_1+a_2)}{16\pi}[(\Lambda_{A1}^2+m^2)^{3/2}-m^3]+\frac{-(b_1+b_2)}{16\pi}\left[(\Lambda_{B1}^2+m^2)^{3/2}-m^3\right].
\end{eqnarray}
Adjusting the cutoffs as $\frac{(\Lambda_{B1}^2+m^2)^{3/2}-m^3}{(\Lambda_{A1}^2+m^2)^{3/2}-m^3}=\frac{a_1+a_2}{b_1+b_2}$ in Eq.\,\eqref{J1(x)-f1} leads to the removal of all infinities of $\mathcal{J}_1$ for the inner regions. The last divergent term indicated in Eq.\,\eqref{After.Second.APSF.Zero.order} is $\mathcal{J}_2(x,y)$. For this term we have:
\begin{eqnarray}\label{J2(x.y)}
      &&\mathcal{J}_2(a_1,a_2)+2\mathcal{J}_2\big(\frac{R-a_1}{2},\frac{R+a_2}{2}\big)+2\mathcal{J}_2\big(\frac{R-a_2}{2},\frac{R+a_1}{2}\big)-\{a_1\to b_1,a_2\to b_2\}\nonumber\\
      &&\hspace{1cm}=\left[a_1a_2+2\frac{R-a_1}{2}\frac{R+a_2}{2}+\frac{R-a_2}{2}\frac{R+a_1}{2}-\{a_1\to b_1,a_2\to b_2\}\right]
      \left(\frac{1}{4\pi}\int_{0}^{\infty}r^2\sqrt{r^2+m^2}dr\right)=0,
\end{eqnarray}
where $r^2=\big(\frac{x\pi}{a_1}\big)^2+\big(\frac{y\pi}{a_2}\big)^2+k_z^2+m^2$.  Therefore, all terms remaining in Eq.\,\eqref{After.Second.APSF.Zero.order} are the Branch-cut terms and they are finite. At the last step, using the BSS defined in Eq.\,\eqref{BSS.CAS.Def.}, the limits $R/b\to\infty$ and $b/a\to\infty$ should be applied. After calculating these limits, we obtained the final expression for the leading order Casimir energy of massive scalar field confined with Dirichlet boundary condition in an open-ended rectangular waveguide with cross-section $a_1\times a_2$ per unit length:
\begin{eqnarray}
       E_{\mbox{\small{Cas.}}}^{(1)}=\Big[\mathcal{B}_1(a_1)+\mathcal{B}_2(a_1,a_2)+\mathcal{B}_3(a_1,a_2)\Big]+\{a_1\leftrightarrow a_2\}
\end{eqnarray}
This result is in agreement with those reported in\,\cite{wolfram.}. This consistency for the zero-order Casimir energy, could enhance the reliability of the introduced configurations for the BSS. In other words, additional lines in the outer regions of Fig.\,(\ref{fig.2}) do not leave any contribution on the Casimir energy values after calculating the limit of $R\to\infty$.

\section{The Determination of Values For Cutoffs}
\label{appendix.CR.technique}
By substituting the result of the integral given in Eq.\,\eqref{cal.of.integration.} in Eq.\,\eqref{VacuumEn.after.firstintegrations.}, we have:
\begin{eqnarray}\label{substitude.cutoff}
     E_{A1}^{(1)}=\frac{-\lambda}{32\pi^2}\sum_{n_1,n'_1=1}^{\infty}\sum_{n_2,n'_2=1}^{\infty}
      &&\hspace{-0.2cm}\left[\frac{1}{a_1a_2}\ln\frac{\Lambda_{A1}}{\omega_{A1}a_1}\ln\frac{\Lambda_{A1}}{\omega'_{A1}a_1}
      +\mathcal{O}\left(\frac{1}{\Lambda_{A1}}\right)\right]
  \nonumber\\
  &&\times\bigg[1+\frac{1}{2}\delta_{n_1,{n'}_1}+\frac{1}{2}\delta_{n_2,{n'}_2}+\frac{1}{4}\delta_{n_1,{n'}_1}\delta_{n_2,{n'}_2}\bigg]+\{a_1\leftrightarrow a_2\},
\end{eqnarray}
where $\omega_{A1}^2=k_{A1}^2+m^2$ and ${\omega'}_{A1}^2={k'}_{A1}^2+m^2$. As expressed in Eq.\,\eqref{substitude.cutoff}, we choose a separate cutoff like $\Lambda_{A1}$ for the other regions. Then, using Eq.\,\eqref{subtraction.vacuums.}, we subtract the vacuum energy of regions from each other as follows:
\begin{eqnarray}\label{subtraction.after.cutoff.adjust.}
     \Delta E^{(1)}_{\mbox{\tiny Vac.}}&=& E^{(1)}_{A'}-E^{(1)}_{B'}
     =E^{(1)}_{A1}+...-E^{(1)}_{B1}-...\nonumber\\
     &&\hspace{-1cm}=\frac{-\lambda}{32\pi^2}\sum_{n_1,n'_1=1}^{\infty}\sum_{n_2,n'_2=1}^{\infty}
     \Bigg[\underbrace{\frac{1}{a_1a_2}\ln\frac{\Lambda_{A1}}{\omega_{A1}a_1}\ln\frac{\Lambda_{A1}}{\omega'_{A1}a_1}}_{\mathcal{H}(a_1,a_2;\Lambda_{A1})}
     +\mathcal{O}\left(\frac{1}{\Lambda_{A1}}\right)
     +...-\underbrace{\frac{1}{b_1b_2}\ln\frac{\Lambda_{B1}}{\omega_{B1}b_1}\ln\frac{\Lambda_{B1}}{\omega'_{B1}b_1}}_{\mathcal{H}(b_1,b_2;\Lambda_{B1})}+\mathcal{O}\left(\frac{1}{\Lambda_{B1}}\right)-...\Bigg]
     \nonumber\\&&\hspace{2.5cm}\times\bigg[1+\frac{1}{2}\delta_{n_1,{n'}_1}+\frac{1}{2}\delta_{n_2,{n'}_2}+\frac{1}{4}\delta_{n_1,{n'}_1}\delta_{n_2,{n'}_2}\bigg]
     +\{a_1\leftrightarrow a_2,b_1\leftrightarrow b_2\},
\end{eqnarray}
In the limit $\Lambda_{A1},\Lambda_{B1}\to\infty$, the indicated function in the above equation $\mathcal{H}(x,y;\Lambda)$ is divergent. To remove the infinite parts of this function in the subtraction process defined by BSS, we should adjust a proper value for cutoffs. To do so, we expand the expression of $\mathcal{H}(x,y;\Lambda)$ as:
\begin{eqnarray}\label{subtracting.H(x,y)}
      \mathcal{H}(a_1,a_2;\Lambda_{A1})-\mathcal{H}(b_1,b_2;\Lambda_{B1})=&&\hspace{-0.5cm}\frac{1}{a_1a_2}\Big[\ln^2\Lambda_{A1}
      -\ln\Lambda_{A1}\ln(\omega_{A1}a_1)-\ln\Lambda_{A1}\ln(\omega'_{A1}a_1)+\ln(\omega_{A1}a_1)\ln(\omega'_{A1}a_1)\Big]\nonumber\\
      &-&\{a_1\to b_1,a_2\to b_2,\Lambda_{A1}\to \Lambda_{B1}\}.
\end{eqnarray}
This expansion helps to manifest the infinite parts of $\mathcal{H}(x,y;\Lambda)$. Now, by choosing the cutoff values based on the following relation, one can be sure that all infinite terms are eliminated:
\begin{eqnarray}\label{adjust.cutoff}
       \ln\Lambda_{B1}\Big[\ln\Lambda_{B1}-(\omega_{B1}+\omega'_{B1})\Big]=\frac{b_1b_2}{a_1a_2}\ln\Lambda_{A1}\Big[\ln\Lambda_{A1}-(\omega_{A1}+\omega'_{A1})\Big].
\end{eqnarray}
We maintain that, in using of BSS supplemented by CR technique, sufficient degrees of freedom in choosing the proper value for cutoffs in each case are always available. Therefore, there is no concern about the value of cutoffs. Similar to what occurred above for infinite terms originated from regions $A1$ and $B1$, it can be conducted for two pair of outer waveguides. The only finite contribution from this subtraction was written in Eq.\,\eqref{All.Four.Terms}.

\section{Calculation of The Casimir Energy For Outer Regions}
\label{appendix.outer.region}
A very important point to mention is that the remaining finite contribution to the Casimir energy coming from the outer waveguides, even after the BSS, is nonzero for finite values of the dimensions of the waveguides. However, as shown in the following, in the limit of large $R$, there is a partial cancellation between those terms, and the remaining terms tend to zero in the limit $R\to\infty$. The finite contribution to the Casimir energy coming from the outer waveguides is:
\begin{eqnarray}\label{Outer.Cas.1}
     E^{(1)\mbox{\tiny Outer}}_{\mbox{\tiny {Cas.}}}&=&\bigg\{\frac{-\lambda}{16\pi^2(R-a_1) (R+a_2)}\bigg[\Big[B_2\big(\frac{R+a_2}{2},\frac{R-a_1}{2}\big)+\ln\big(\frac{m(R-a_1)}{2}\big)-m\Big(\frac{R-a_1}{2}+\frac{R+a_2}{2}\Big)\nonumber\\
     &+&\frac{m^2(R-a_1)(R+a_2)}{8\pi}(1+\ln4)
     -\underbrace{\ln(1-e^{-m(R-a_1)})}_{\mathcal{K}_1}\Big]B_2\big(\frac{R+a_2}{2},\frac{R-a_1}{2}\big)\nonumber\\
     &+&B_1\big(\frac{R-a_1}{2}\big)\Big[\frac{m^2(R-a_1)(R+a_2)}{16\pi}(1+\ln4)+\ln \big(\frac{m(R-a_1)}{2}\big)-\frac{m(R+a_2)}{4}\Big]\nonumber\\
     &+&B_1\big(\frac{R+a_2}{2}\big)\Big[\frac{m(R+a_2)}{2}-B_1\big(\frac{R+a_2}{2}\big)\Big] +\underbrace{\frac{m(R-a_1)}{2}\int_{1}^{\infty}d\eta\frac{\ln(\eta^2-1)}{e^{m(R-a_1)\eta}-1}}_{\mathcal{K}_2}
     \nonumber\\
     &+&I_8\big(\frac{R-a_1}{2},\frac{R+a_2}{2}\big)+I_9\big(\frac{R-a_1}{2},\frac{R+a_2}{2}\big)+I_{10}\big(\frac{R-a_1}{2},\frac{R+a_2}{2}\big)
     +B_2\big(\frac{R-a_1}{2},\frac{R+a_2}{2}\big)\nonumber\\
     &-&\underbrace{\frac{m^2(R-a_1)(R+a_2)}{2\pi}\int_{1}^{\infty}d\eta\frac{\sqrt{\eta^2-1}\ln(\eta^2-1)}{e^{m(R+a_2)\eta}-1}}_{\mathcal{K}_3}
     -\underbrace{\frac{m(R+a_2)}{2}\int_{\epsilon}^{1}dN\frac{\ln\Big(1-e^{-m(R-a_1)\sqrt{1-N^2}}\Big)}{e^{m(R+a_2)N}-1}}_{\mathcal{K}_4}\nonumber\\
     &-&\frac{1}{2}\ln[1-e^{-m(R-a_1)}]\ln\epsilon\bigg]
     +\{a_1\leftrightarrow a_2\}\bigg\}-\{a_1\to b_1,a_2\to b_2\},
\end{eqnarray}
All terms in the above expression should be evaluated in the limit $R\to\infty$. To do so, we first compute the Taylor expansion of the following term:
\begin{eqnarray}\label{Taylor.expansion.1}
      B_2\big(\frac{R+a_2}{2},\frac{R-a_1}{2}\big)
      =\frac{-m(R+a_2)}{2\pi}\sum_{j=1}^{\infty}\frac{K_1(mj(R-a_1))}{j}\buildrel{R\to\infty}\over\longrightarrow
      \frac{-m\sqrt{R}}{2\sqrt{2\pi m}}\sum_{j=1}^{\infty}\frac{e^{-mj(R-a_1)}}{j\sqrt{j}}+\mathcal{O}(1/\sqrt{R}).
\end{eqnarray}
The expressions $I_8\big(\frac{R-a_1}{2},\frac{R+a_2}{2}\big)$, $I_9\big(\frac{R-a_1}{2},\frac{R+a_2}{2}\big)$, $I_{10}\big(\frac{R-a_1}{2},\frac{R+a_2}{2}\big)$, and $B_1\big(\frac{R-a_1}{2}\big)$ tend to zero when the limit $R\to\infty$. Moreover, all terms indicated by $\mathcal{K}_1$, $\mathcal{K}_2$, $\mathcal{K}_3$, and $\mathcal{K}_4$ in Eq.\,\eqref{Outer.Cas.1} diminish when the limit $R\to\infty$ is applied. So, Eq.\,\eqref{Outer.Cas.1} is converted to:
\begin{eqnarray}\label{Outer.Cas.2}
    E^{(1)\mbox{\tiny Outer}}_{\mbox{\tiny {Cas.}}}&\approx&\bigg\{\frac{-\lambda}{16\pi^2(R-a_1) (R+a_2)}\bigg[\Big[\frac{-m\sqrt{R}}{2\sqrt{2\pi m}}\sum_{j=1}^{\infty}\frac{e^{-mj(R-a_1)}}{j\sqrt{j}}+\mathcal{O}(1/\sqrt{R})+\ln\big(\frac{m(R-a_1)}{2}\big)\nonumber\\
     &-&m\Big(\frac{R-a_1}{2}+\frac{R+a_2}{2}\Big)+\frac{m^2(R-a_1)(R+a_2)}{8\pi}(1+\ln4)\Big]\Big[\frac{-m\sqrt{R}}{2\sqrt{2\pi m}}\sum_{j=1}^{\infty}\frac{e^{-mj(R-a_1)}}{j\sqrt{j}}+\mathcal{O}(1/\sqrt{R})\Big]\nonumber\\
     &+&\frac{-m\sqrt{R}}{2\sqrt{2\pi m}}\sum_{j=1}^{\infty}\frac{e^{-mj(R+a_2)}}{j\sqrt{j}}+\mathcal{O}(1/\sqrt{R})\bigg]
     +\{a_1\leftrightarrow a_2\}\bigg\}-\{a_1\to b_1,a_2\to b_2\},
\end{eqnarray}
The limit $R\to\infty$ in the above expression yields all terms go to zero. Therefore, the outer waveguides have done
their job in the BSS of cancelling infinities without having any finite effect on the Casimir energy in the limit $R\to\infty$.

\acknowledgments
The Author would like to thank the research office of Semnan Branch, Islamic Azad University for the financial support.

\end{document}